\documentclass{PoS}

\def\ergscm2{erg\,s$^{-1}$cm$^{-2}$}

\def\uu{4U\,0142+614\,}
\def\ee{1E\,1048-5937\,}
\def\kes{1E\,1841-045\,}

\def\rxs{1RXS\,J1708-4009\,}
\def\ea{1E\,2259+586\,}
\def\xte{XTE\,J1810-197\,}

\def\wes{CXOU\,J1647-4552\,}
\def\1e{1E\,1547.0-5408\,}
\def\sgra{SGR\,1806-20\,}
\def\sgrb{SGR\,1900+14\,}

\newcommand{\XMM}{{\it XMM--Newton}\,}

\newcommand{\RXTE}{{\it R}XTE\,}
\newcommand{\INT}{{\it INTEGRAL}\,}

\usepackage{lscape}

\title{Modelling magnetars' high energy emission through Resonant Cyclotron Scattering}

\ShortTitle{Resonant Cyclotron Scattering}

\author{\speaker{Diego G\"otz}\\
        CEA Saclay - DSM/Irfu/Service d'Astrophysique UMR AIM - Orme des Merisiers, Bat. 709, F-91191 Gif-sur-Yvette, France\\
        E-mail: \email{diego.gotz@cea.fr}}

\author{Nanda Rea\\
        University of Amsterdam, Astronomical Institute Anton Pannekoek, Kruislaan, 403, 1098 SJ, Amsterdam, The Netherlands\\}
\author{Silvia Zane\\
        \llap{$^\dagger$}Mullard Space Science Laboratory, University College London, Holmbury St. Mary, Dorking, Surrey, RH5 6NT, UK\\}

\author{Roberto Turolla$^{\dagger}$\\
        Department of Physics, University of Padova, Via Marzolo 8, I-35131 Padova, Italy\\}

\author{Maxim Lyutikov\\
        Department of Physics, Purdue University, 525 Northwestern Avenue, West Lafayette, IN 47907, USA\\}

\abstract{We present a systematic fit of a model of resonant cyclotron scattering (RCS) to the X and soft $\gamma$-ray data of four magnetars, including anomalous X-ray pulsars, and soft gamma repeaters. In this scenario, non-thermal magnetar spectra in the soft X-rays result from resonant cyclotron scattering of the thermal surface emission by hot magnetospheric plasma. We find that this model can successfully account for the soft X-ray emission of magnetars, while an additional component is still required to model the hard X-ray persistent magnetar emission recently discovered by \INT. The latter is an important component in terms of magnetars' luminosity, and cannot be neglected when modelling the soft X-ray part of the spectrum.}

\FullConference{7th INTEGRAL Workshop\\
		 September 8-11 2008\\
		 Copenhagen, Denmark}

\begin{document}

\section{Introduction}
Soft Gamma Repeaters (SGRs) and Anomalous X-ray Pulsars (AXPs) (see \cite{mereghetti08} for a recent rewiev) are
believed to be Magnetar candidates, namely isolated neutron stars powered by the decay of a huge ($B > B_{crit}\simeq 4.4\times 10^{13} \; \rm G$) magnetic field \cite{dt92}. They are slow rotators ($P\sim$ 2--12 s), rapidly spinning down
($\dot P\sim$10$^{-(11-12)}$ s s$^{-1}$), with rather soft (at least for AXPs) X-ray spectra that can be modelled by
the sum of a black body ($kT\sim$ 0.5 keV) and a power law component ($\Gamma\sim$ 2--4). Some of them are
transients, and they sporadically emit short ($\sim$0.1 s) bursts of high-energy radiation during unpredictable 
periods of activity. SGRs also emit giant flares: very powerful short spikes ($E > 10^{44}$ erg), followed by pulsating
tails lasting a few hundreds of seconds. On the other hand the emission of short bursts by AXPs is sometimes associated
with large flares in the persistent emission lasting hundreds of days, when the flux can vary over more than two orders
of magnitudes.

As mentioned before, X-ray (0.1--10 keV) spectra of AXPs and SGRs are well described by an empirical model composed
by the sum of a black body and a power law modified by absorbtion. But as soon as the first persistent counterparts
of magnetars have been reported at high energies ($>$20 keV) by the \INT\ observations (e.g. \cite{dg06}), it was
evident that this simple model was not adapted to describe the broad-band persistent spectra of Magnetars. Indeed,
the \INT\ data showed the presence in five sources of hard tails up to $\sim$ 200 keV which lie well above the extrapolation of the X-ray data, especially for AXPs. This shows that at least some magnetars emit as much as, or even more, energy at hard X-rays than below 10 keV. Stimulated by this, and by the necessity of giving a physical interpretation
of the two components at low energies, we implemented a simplified Resonant Cyclotron Scattering model
and we fitted all magnetar spectra for which good quality \XMM\ and \INT\ data are available. Our detailed
results are reported in \cite{rea08}. Here we focus only on the sources for which \INT\ data are available. 

\section{Resonant Cyclotron Scattering}
\label{rcsmodel}
Before discussing our {\tt XSPEC} model and the implications of our
results, we briefly touch on some properties of the RCS model which
directly bear to the physical interpretation of the fitting parameters
and their comparison with similar parameters introduced in other
theoretical models. The basic idea follows the original suggestion by
\cite{tlk} (hereafter TLK), who pointed out that a scattering plasma may be supplied to the
magnetosphere by plastic deformations of the crust, which twist the
external magnetic field and push electric currents into the
magnetosphere. The particle density of charge carries required to
support these currents may largely exceed the Goldreich-Julian charge
density \cite{Goldandjul69}. Furthermore, it is expected that
instabilities heat the plasma.

Following this idea, \cite{Lyutikov06} studied how
magnetospheric plasma might distort the thermal X-ray emission
emerging from the star surface through efficient resonant cyclotron
scattering.  If a large volume of the neutron star magnetosphere is
filled by a hot plasma, the thermal (or quasi-thermal) cooling radiation
emerging from the star surface will experience repeated scatterings at
the cyclotron resonance.
The efficiency of the process is quantified by the scattering optical
depth, $\tau_{res}$,

\begin{equation}
\tau_{res}= \int \sigma_{res} \; n_e \; dl = \tau_0 ( 1+ \cos^2 \alpha)
\label{eqtaures}
\end{equation}
where

\begin{equation}
\sigma_{res} = {\sigma_T \over 4} { (1+ \cos ^2 \alpha) \omega^2 \over (
\omega - \omega_B)^2 +\Gamma^2/4}
\label{sigres}
\end{equation}
is the (non-relativistic) cross-section for electron scattering in the
magnetized regime, $n_e$ is the electron number density, $\alpha$ is the
angle between the photon propagation
direction and the local magnetic field,
$\Gamma = 4 e^2 \omega_B^2 / 3 m_e c^3$ is the natural width of the first
cyclotron harmonic, $\sigma_T$ is the
Thomson scattering cross-section, and
\begin{equation}
\tau_0 = { \pi^2 e^2 n_e r\over 3 m_e c \omega_B} \, .
\label{tau0}
\end{equation}
Here $r$ is the radial distance from the center of the star,
$\omega_B = eB/m_{e}c$ is the electron cyclotron frequency, and $B$ is
the local value of the magnetic field. At energies corresponding to soft
\mbox{X-ray} photons, the resonant scattering optical depth greatly exceeds that
for Thomson scattering, $\tau_{T} \sim n_e \sigma_T r$,
\begin{equation}
 { \tau_{res} \over \tau_{T} } \sim { \pi \over 8} { m_ec^3 \over e^2
\omega_B} \sim 10^5 \left( \frac{1\ \mathrm{keV}}{\hbar \omega_B}
\right)\,.
\end{equation}
This implies that even a relatively small amount of plasma
present in the magnetosphere of the NS may considerably modify the
emergent spectrum.

The RCS model developed by \cite{Lyutikov06}, and used in
this investigation, is based on a simplified, 1D semi-analytical
treatment of resonant cyclotron up-scattering of soft thermal photons,
under the assumption that scattering occurs in a static,
non-relativistic, warm medium and neglecting electron recoil. The
latter condition requires $\hbar\omega\ll m_ec^2$.  Emission from the
neutron star surface is treated assuming a blackbody spectrum, and
that seed photons propagate in the radial direction.
Magnetospheric charges are taken to have a top-hat
velocity distribution centered at zero and extending up to
$\pm\beta_T$. Such a velocity distribution mimics a scenario in which
the electron motion is thermal (in 1D because charges stick to the
field lines). In this respect, $\beta_T$ is associated to the mean
particle energy and hence to the temperature of the 1D electron
plasma.
Since scatterings
with the magnetospheric electrons occur in a thin shell of width $H
\sim \beta_{T} r/3 \ll r$
around the ``scattering sphere'', one can treat
the scattering region as a plane-parallel slab. Radiation transport is
tackled by assuming that photons can only propagate along the slab
normal, i.e. either towards or away from the star. Therefore,
$\cos\alpha=\pm 1$ in eq.~(\ref{eqtaures}) and it is $\tau_{res}= 2
\tau_0$; the electron density is assumed to be constant through the
slab. We notice that the model does not account for the bulk
motion of the charges. This is expected since the starting point is
not a self-consistent calculation of the currents but a prescription
for the charge density.  As a consequence, the electron velocity and
the optical depth are independent parameters, although in a more
detailed treatment this might not be the case \cite{belobodorov07}.

Although Thomson scattering conserves the photon energy in the
electron rest frame, the (thermal) motion of the charges induces a
frequency shift in the observer frame. However, since our electron
velocity distribution averages to zero, a photon has the same
probability to undergo up or down-scattering. Still, a net
up-scattering (and in turn the formation of a hard tail in the
spectrum) is expected if the magnetic field is inhomogeneous.  For a
photon propagating from high to low magnetic fields, multiple resonant
cyclotron scattering will, on average, up-scatter in energy the
transmitted radiation, while the dispersion in energy decreases with
optical depth \cite{Lyutikov06}.  Photon boosting by particle
thermal motion in Thomson limit occurs due to the spatial variation of
the magnetic field and differs qualitatively from the (more familiar)
non-resonant Comptonization \cite{kompaneets56}. As a result, the
emerging spectrum is non-thermal and under certain circumstances can
be modeled with two-component spectral models consisting of a
blackbody plus a power-law \cite{Lyutikov06}.

\begin{figure}
\centering{
\includegraphics[height=6cm,width=6cm,angle=-90]{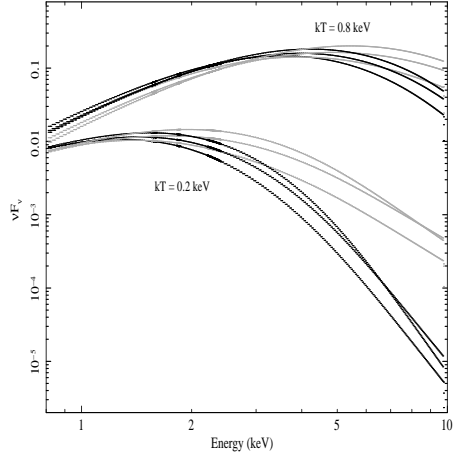}}
\caption{Distorsion of a seed blackbody spectrum through
resonant cyclotron scattering onto magnetosferic electrons, for
two values of the blackbody temperature, 0.2~keV and 0.8~keV. Black
lines show the
RCS model for $\beta_{T}=0.2$ and $\tau_{res} =$ 2, 4, 8 (from bottom to
top), while grey lines are relative to $\beta_{T}=0.4$ and $\tau_{res} =$ 2,
4, 8 (from bottom to top). The normalizations of the various curves are
arbitrary.}
\label{figrcs}
\end{figure}

In order to implement the RCS model in {\tt XSPEC}, we created a grid
of spectral models for a set of values of the three parameters
$\beta_{T}$, $\tau_{res}$ and $T$. The parameter ranges are
$0.1\leq\beta_{T}\leq0.5$ (step 0.1; $\beta_T$ is the thermal velocity
in units of $c$), $1
\leq\tau_{res}\leq 10$ (step 1; $tau_{res}$ is the optical depth) 
and $0.1$~keV$\leq T\leq1.3$~keV (step 0.2\,keV; $T$ is the
temperature of the seed thermal surface radiation, assumed to be a
blackbody). For each model, the spectrum was computed in the energy
range 0.01--10\,keV (bin width 0.05\,keV). The final {\tt XSPEC} {\tt
atable} spectral model has therefore three parameters, plus the
normalization constant, which are simultaneously varied during the
spectral fitting following the standard $\chi^2$ minimization
technique. In Fig.\,\ref{figrcs} we show the comparison between a
blackbody model and our RCS model. We stress again that our model has
the same number of free parameters (three plus the normalization) than
the blackbody plus power-law or two blackbody models ($\beta_{T}$,
$\tau_{res}$, $T$, plus the normalization, compared to $kT$, $\Gamma$
(or $kT_2$), plus two normalizations); it has then the same
statistical significance. We perform in the following section a
quantitative comparison between the RCS model and other models
commonly used in the soft X-ray range. However, note that here the RCS
model is meant to model spectra in the 0.1--10\,keV energy range. For
all sources with strong emission above $\sim 20$~keV, the spectrum was
modeled by adding to the RCS a power-law meant to reproduce the hard
tail (see \S\,\ref{results} for details). This power-law does not have
(yet) a clear physical meaning in our treatment, but since it
contributes also to the 0.1--10\,keV band, our RCS parameters depend
on the correct inclusion of this further component.

\section{Results}
\label{results}
The complete dataset, used in \cite{rea08}, contains a set of AXPs which 
emit in the hard X-ray range, and also happen to be ``steady'' emitters or showing moderate flux and spectral
variability (flux changes less than a factor of 5), and a group with larger
variations: \uu, \rxs, \kes, and \ea.  Then a set of ``transient'' AXPs (often labeled TAXPs), which
includes \xte, \1e, and \wes, was considered. To these we added \ee, in the light of the
recent detection of large outbursts from this source (e.g. \cite{mereghetti04}), 
and of its spectral similarities with canonical TAXPs. Finally, a set of SGRs, which comprises \sgra, and \sgrb.

Here we report in detail only the results of the fits concerning the five magnetars that have been firmly detected with \INT.
For a complete description of the observations, as well as for data analysis techniques, see \cite{rea08}.

\subsection{AXPs: the hard X-ray emitters}
\label{res:hard}

In this section we first consider the AXPs with \INT\ detected hard X-ray
emission, which also coincides with the marginally variable AXPs. 
We recall that, strictly speaking, these hard X-ray emitting
AXPs are not ``steady'' X-ray emitters. Subtle flux and spectral
variability was discovered in \rxs\, and \uu. In particular, \rxs\,
showed a long term, correlated intensity-hardness variability (both in
the soft and hard X-rays), most probably related to its glitching
activity (e.g. \cite{israel07}). \uu\, showed a flux increase
of $\sim 10\%$ (also correlated with a spectral hardening) following
the discovery of its bursting activity \cite{dib07}.  Furthermore, thanks to a large
\RXTE\ monitoring campaign, long-term spin period variations and
glitches were discovered in \uu\, \rxs, and \kes, i.e. the three AXPs
which are the brightest both in the soft and hard X-ray bands \cite{gavriil02,dallosso03,dib07,israel07}.

\begin{figure*}
\centering{
\hspace{0.1cm}
\vbox{
\hbox{
\includegraphics[height=2.5cm,angle=270,width=5.cm]{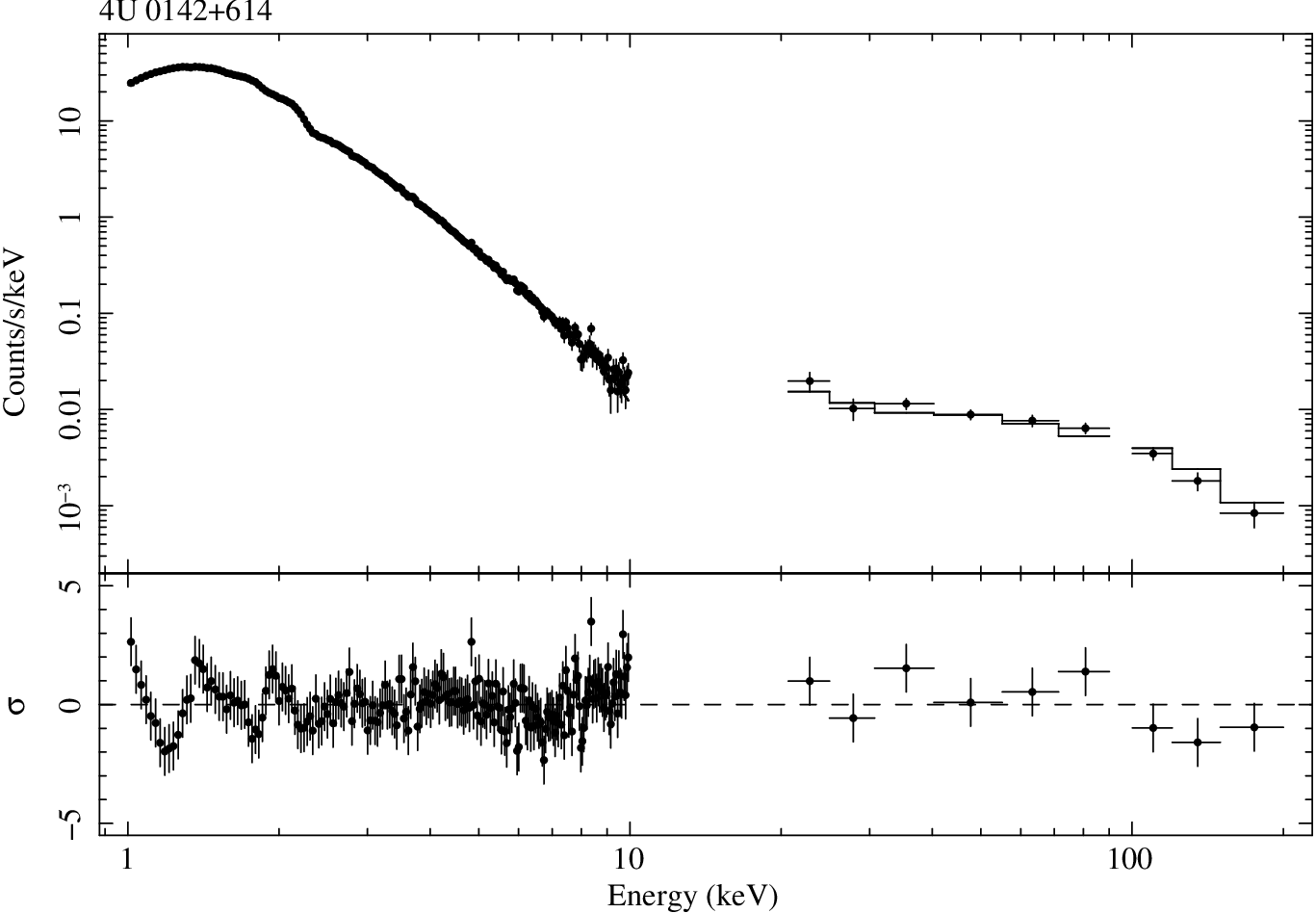}
\includegraphics[height=2.5cm,angle=270,width=5.cm]{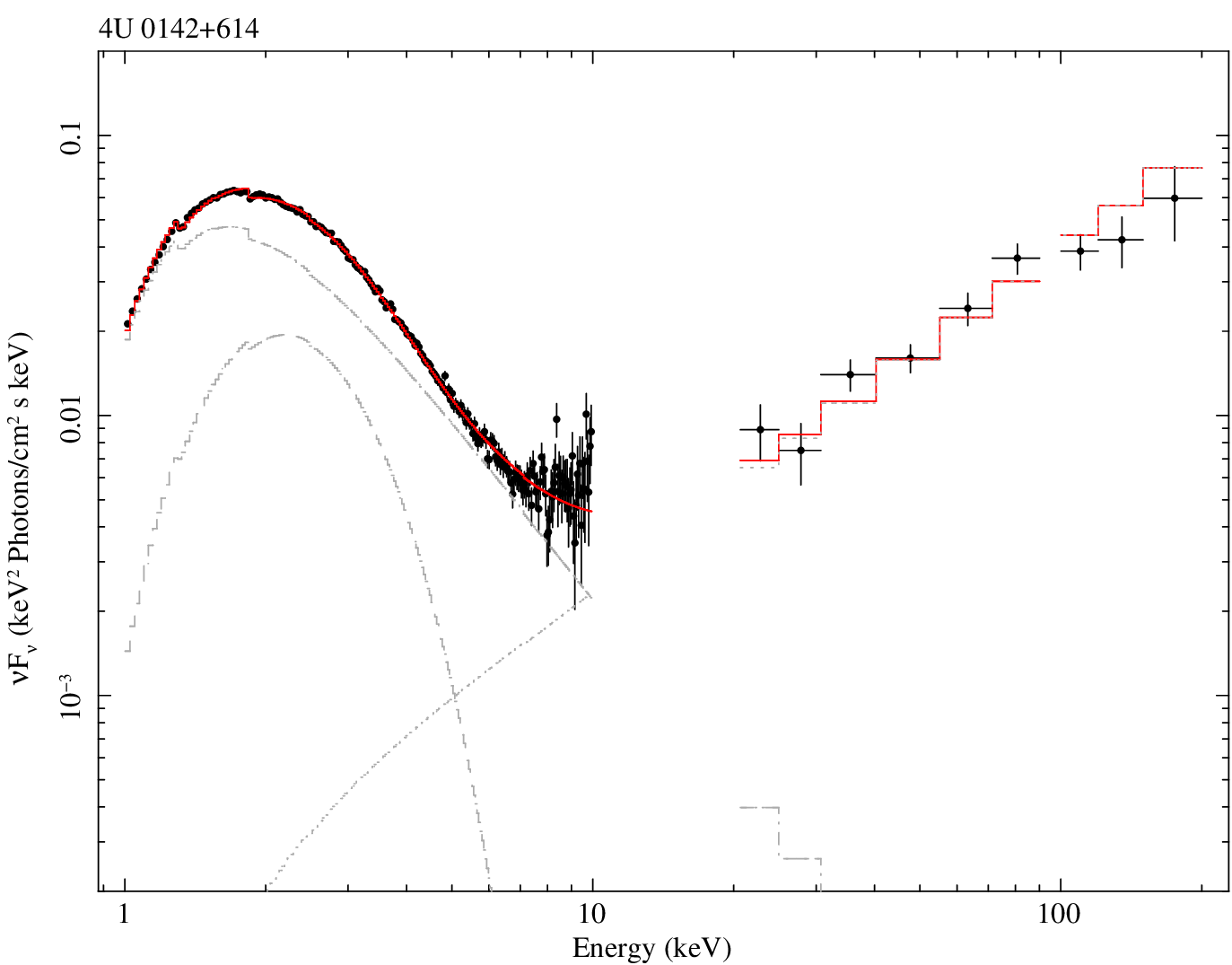}}
\hbox{
\includegraphics[height=2.cm,angle=270,width=5.cm]{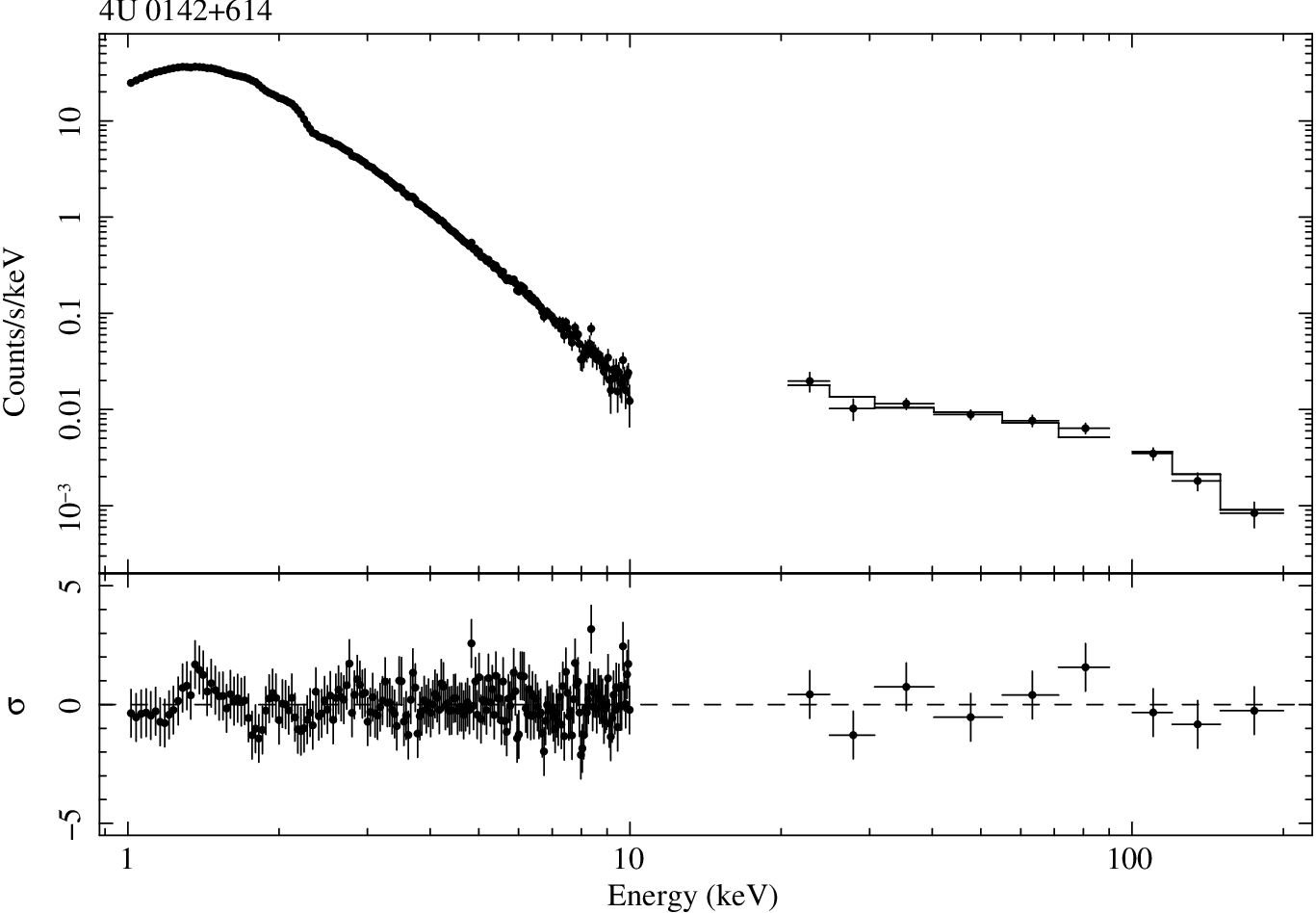}
\includegraphics[height=2.cm,angle=270,width=5.cm]{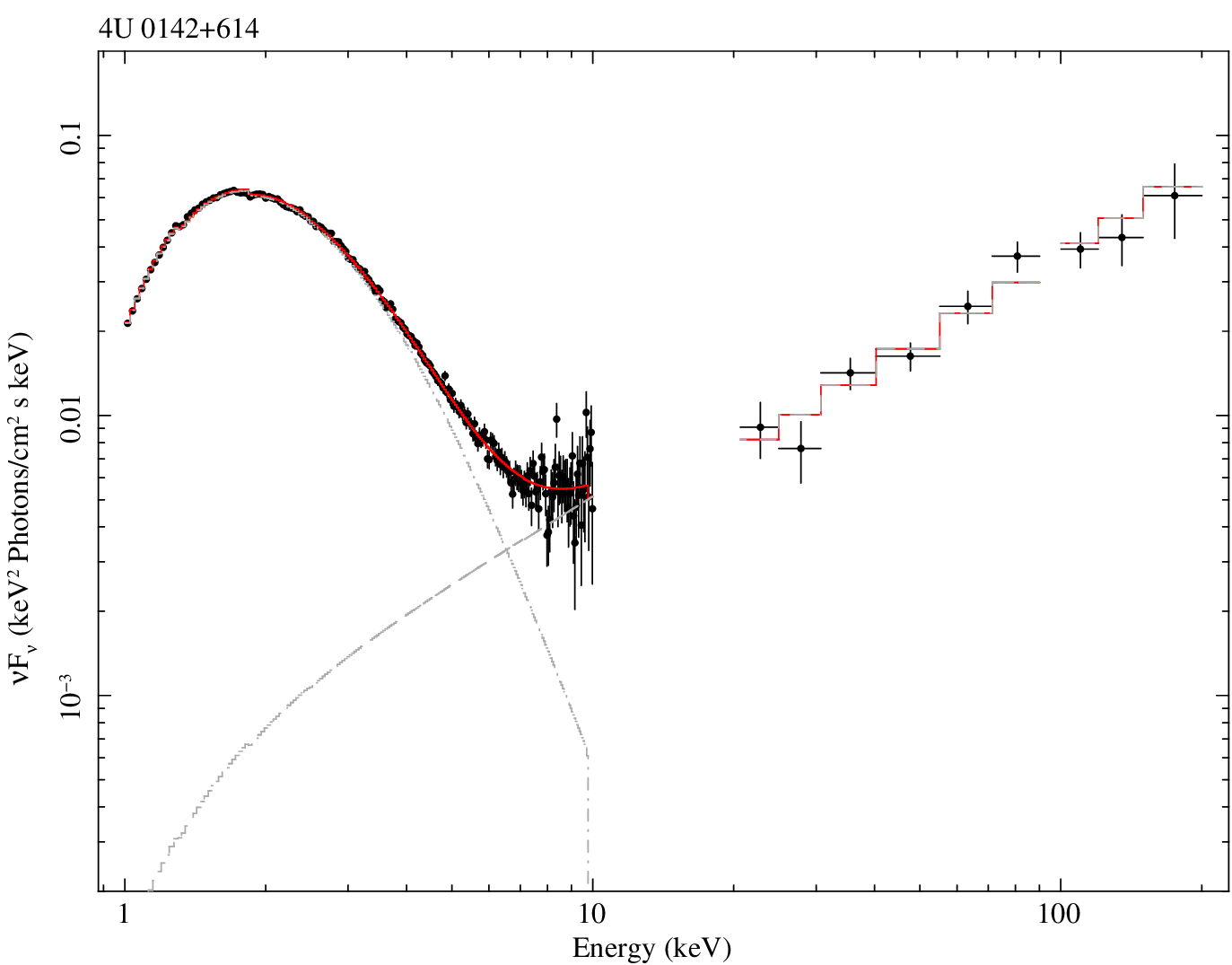}}
\hbox{
\includegraphics[height=2.cm,angle=270,width=5.cm]{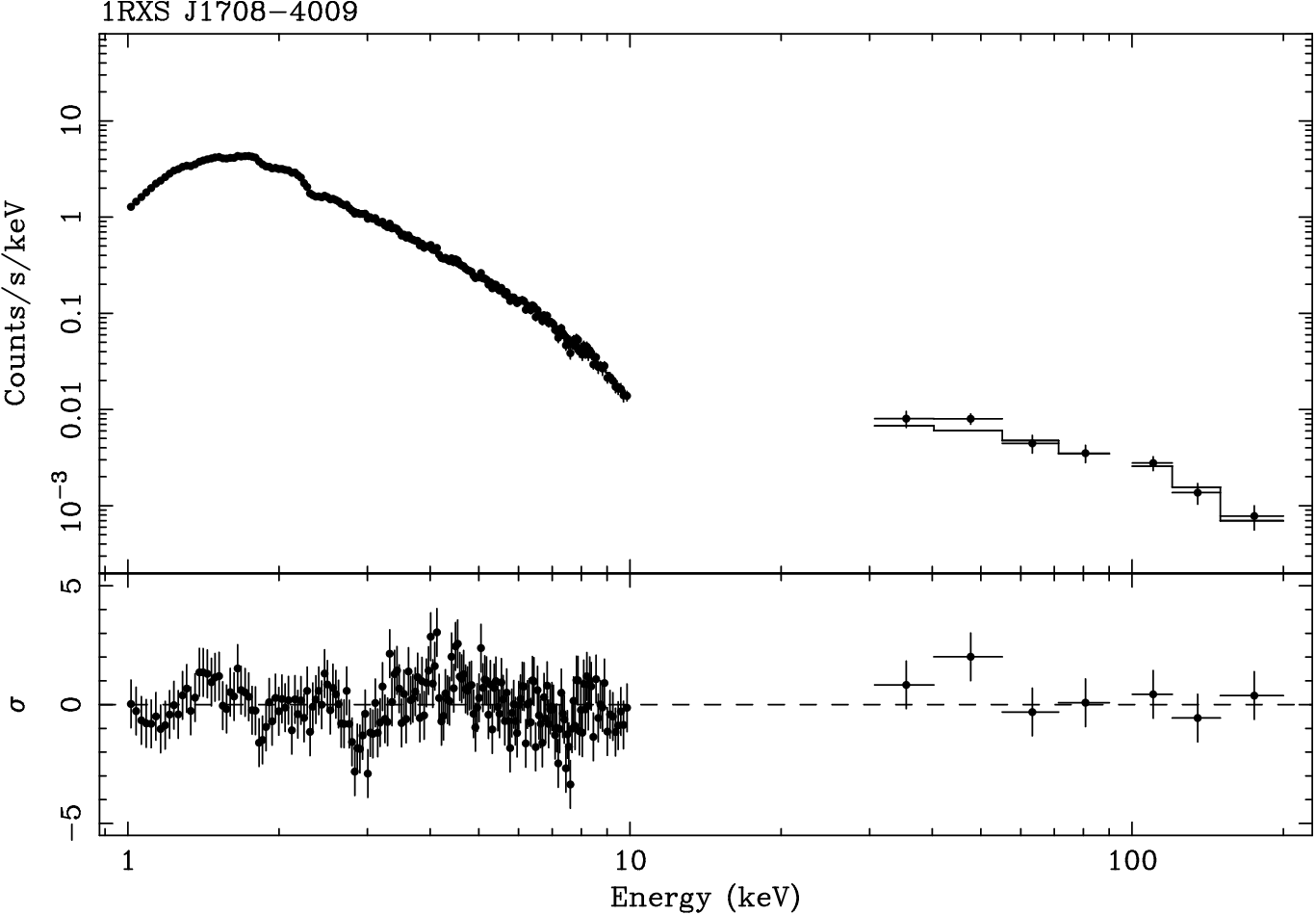}
\includegraphics[height=2.cm,angle=270,width=5.cm]{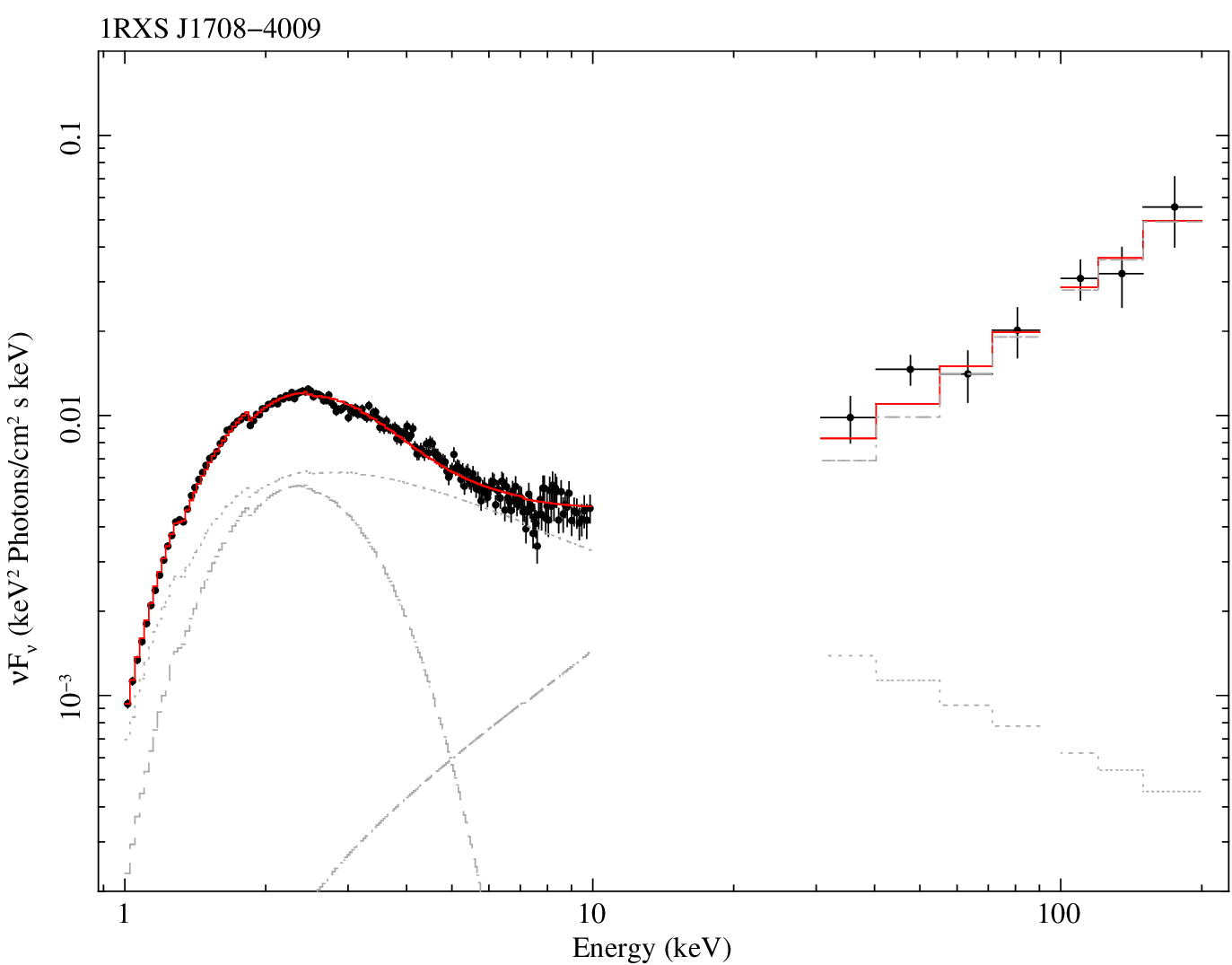}}
\hbox{
\includegraphics[height=2.cm,angle=270,width=5.cm]{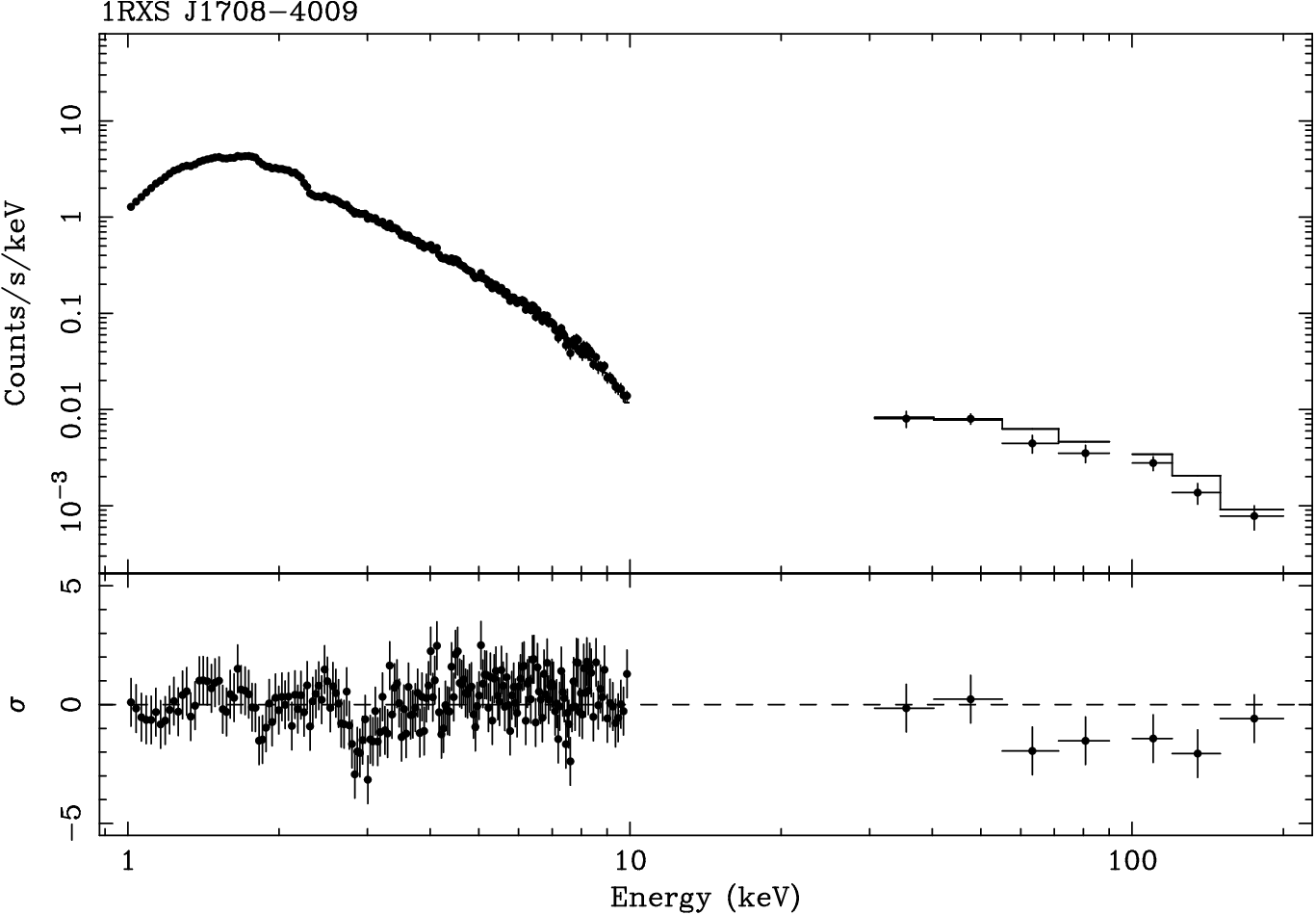}
\includegraphics[height=2.cm,angle=270,width=5.cm]{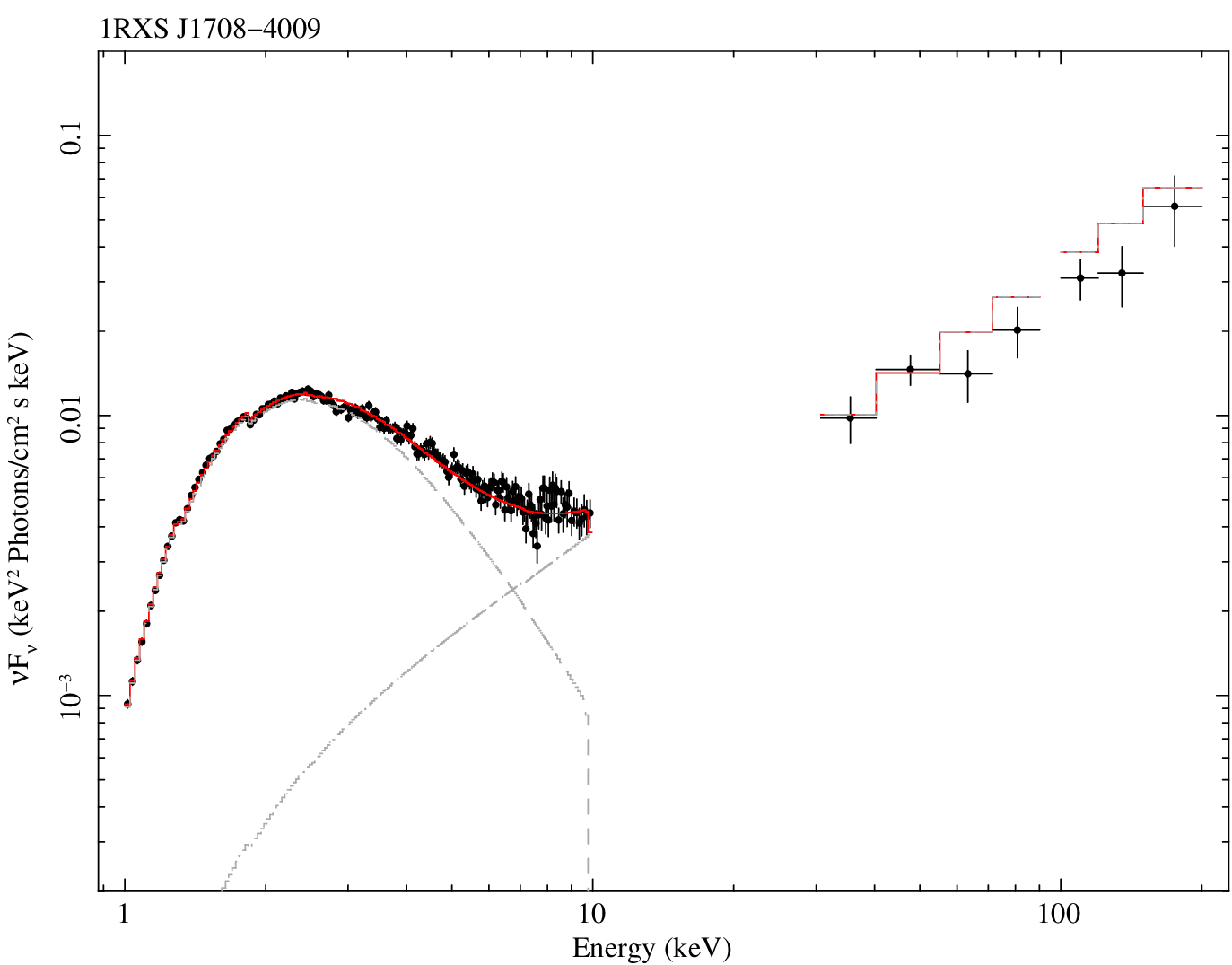}}
}}
\caption{\uu, and \rxs: left column shows the spectra in Counts/s/keV
while in the right column we report the $\nu$F$_\nu$ plots. The upper panels are relative to the modeling with a
blackbody plus two power-laws (BB+2PL). Bottom panels report for both sources the
resonant cyclotron scattering plus a power-law model (RCS+PL).} 
\label{spectraxmmintegral1}
\end{figure*}

\begin{figure}[b!]
\centering{
\hspace{0.1cm}
\vbox{
\hbox{
\includegraphics[height=2.cm,angle=270,width=5.cm]{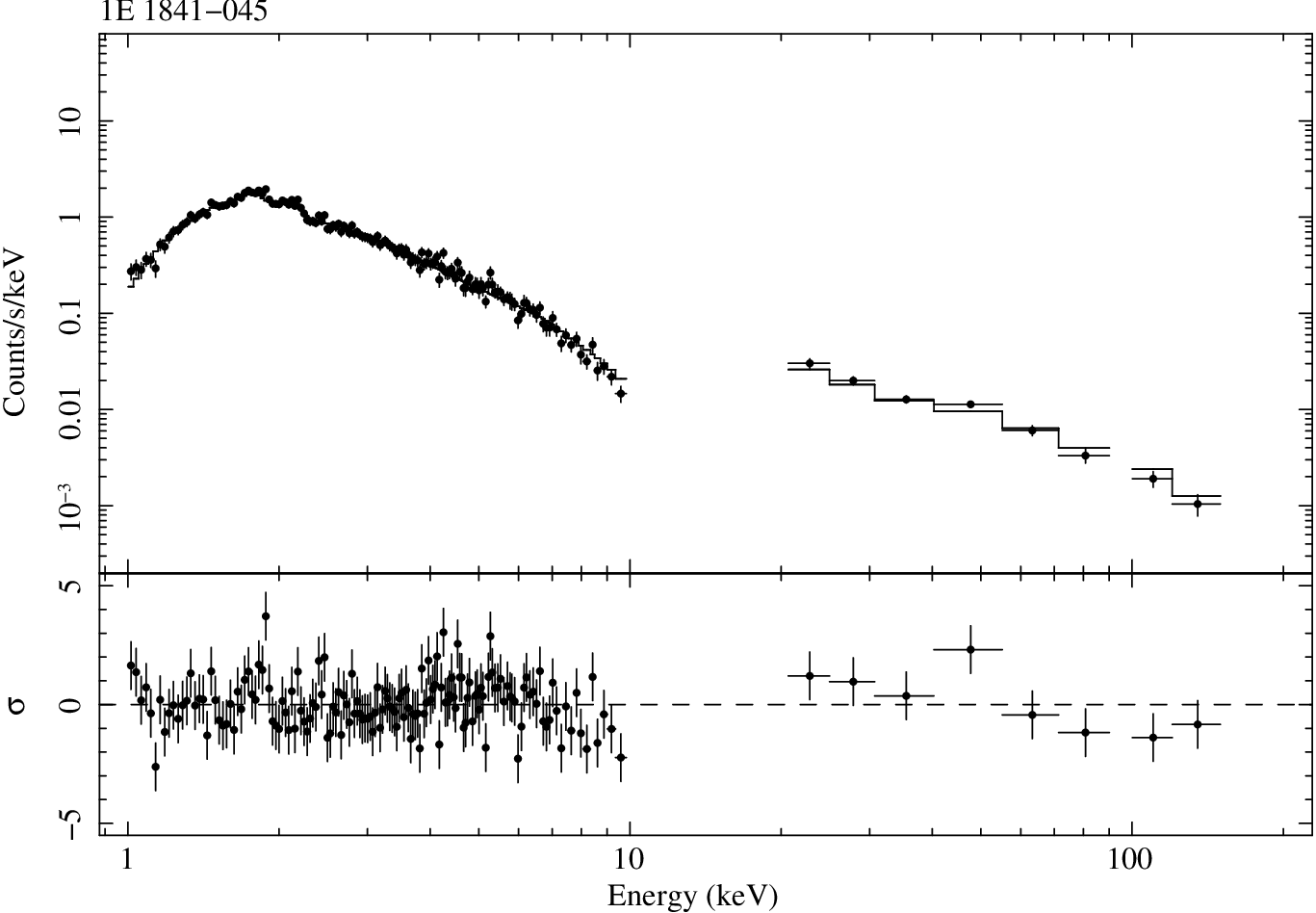}
\includegraphics[height=2.cm,angle=270,width=5.cm]{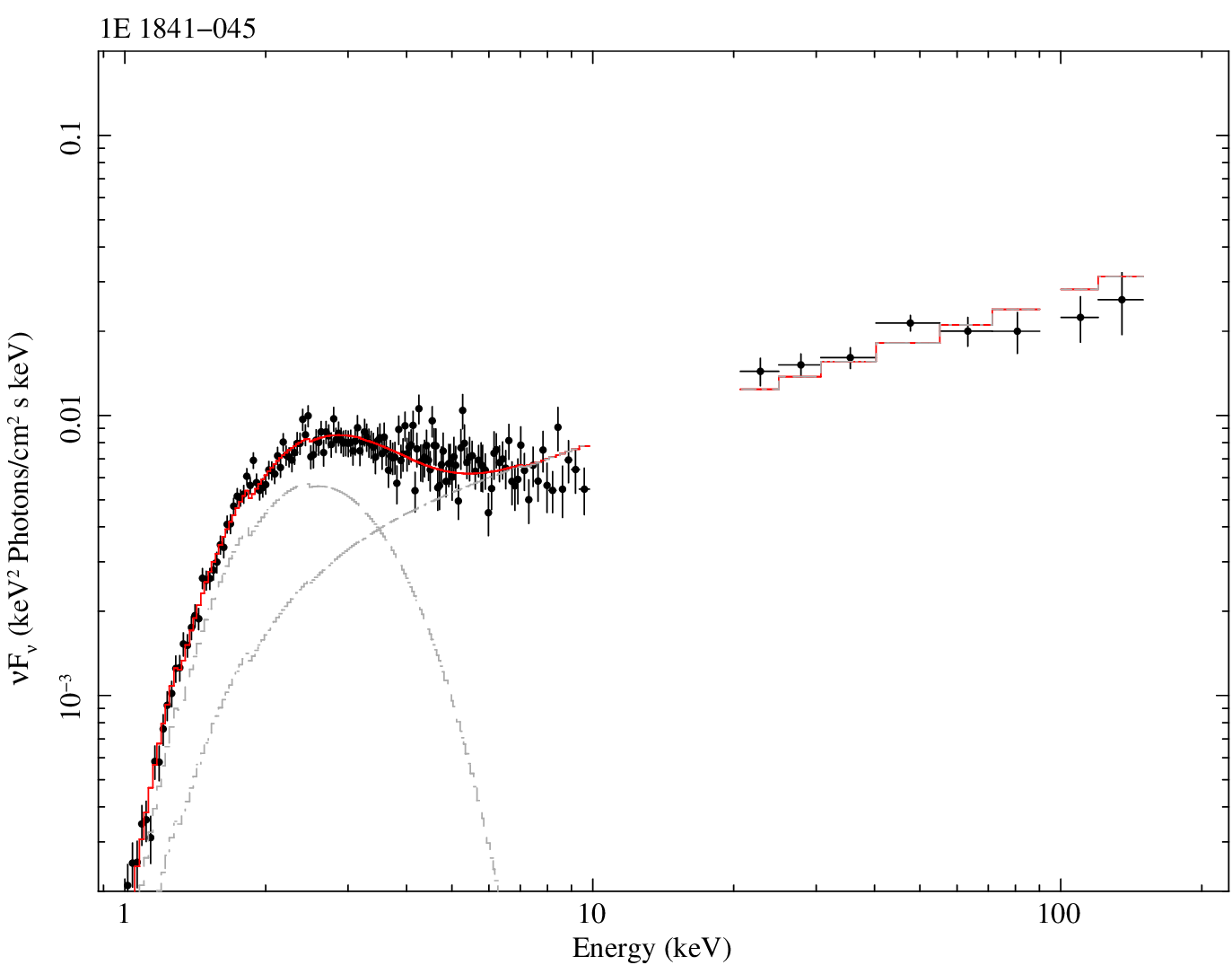}}
\hbox{
\includegraphics[height=2.cm,angle=270,width=5.cm]{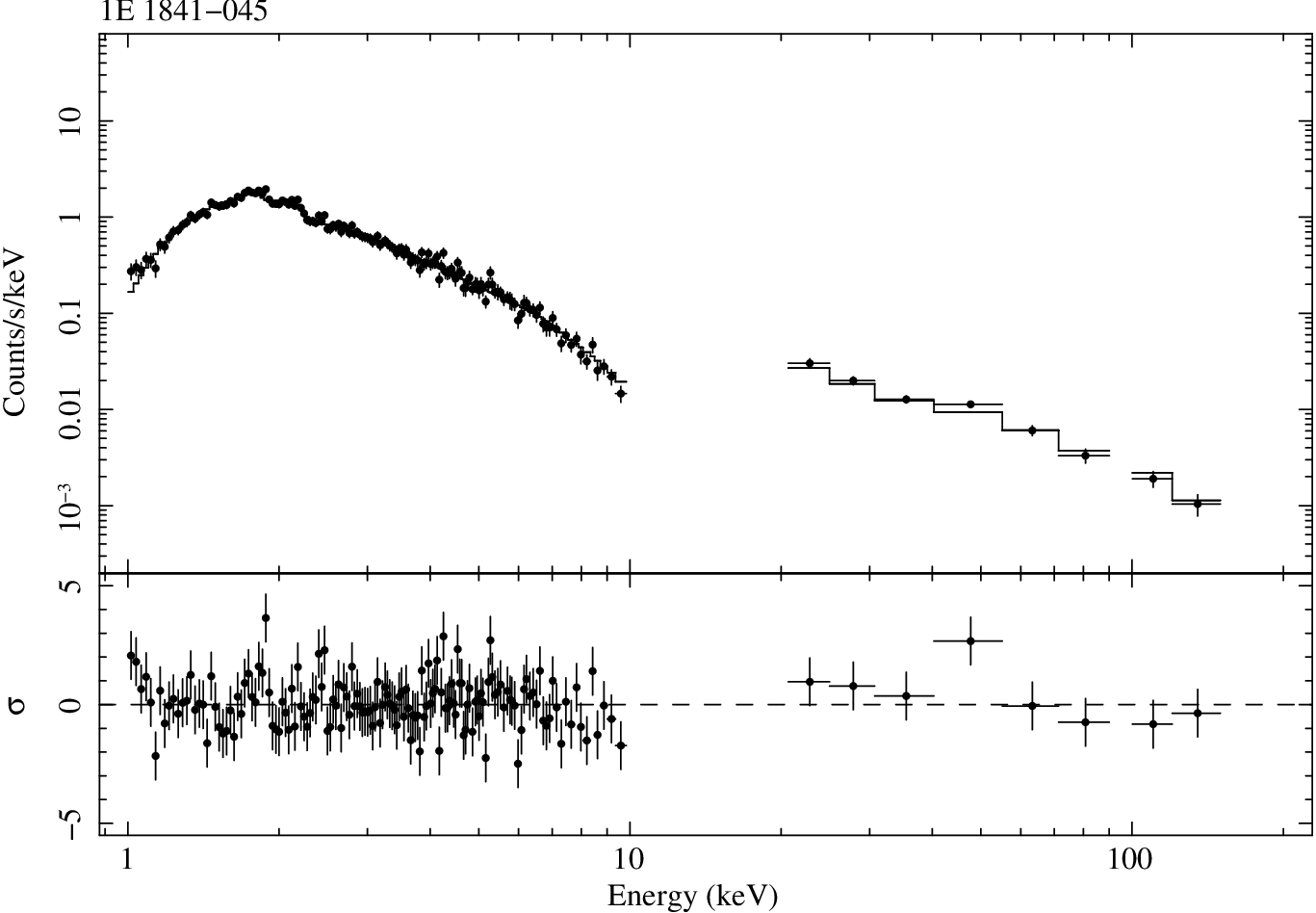}
\includegraphics[height=2.cm,angle=270,width=5.cm]{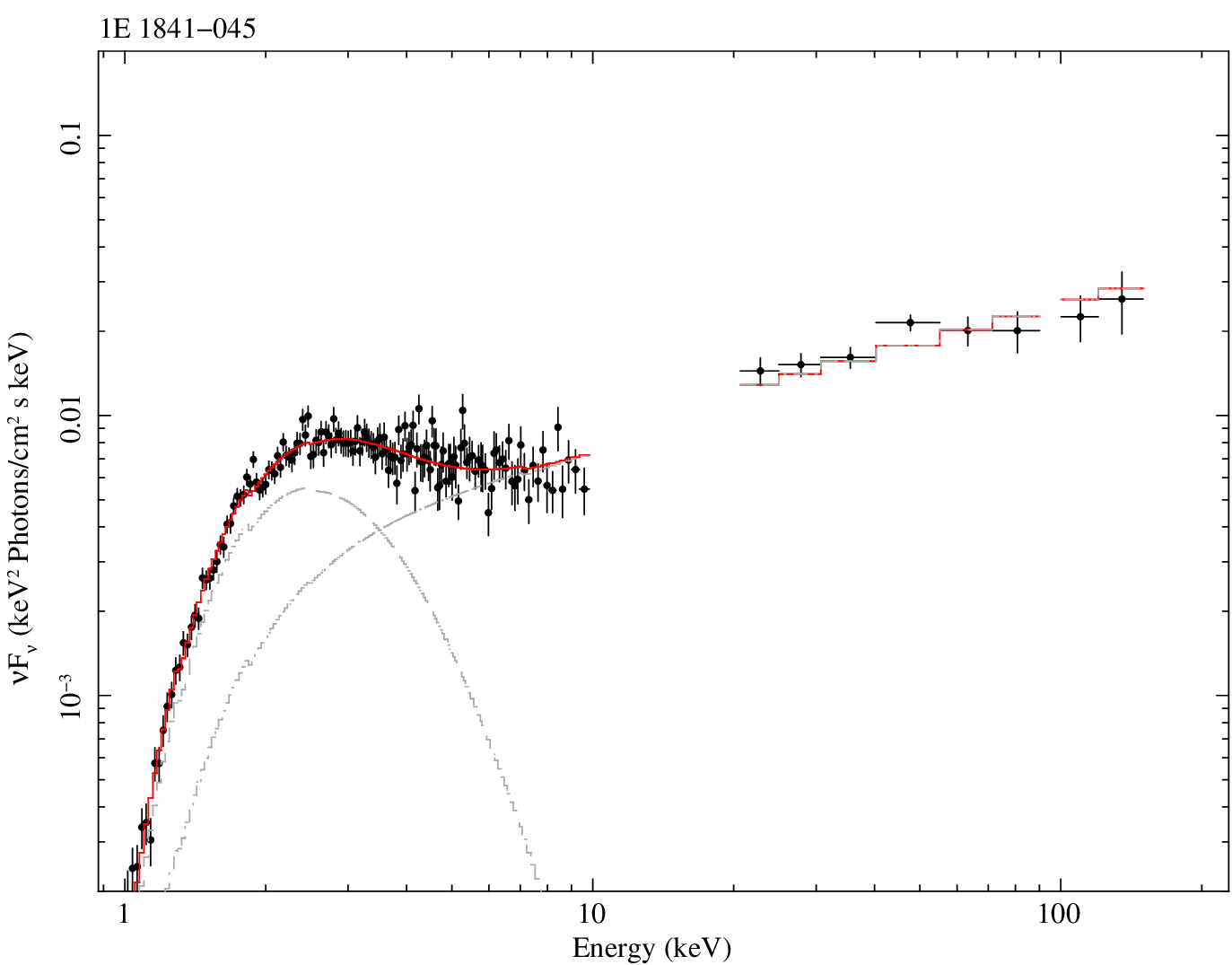}} } }
\caption{\kes: left column shows the spectra in Counts/s/keV
while in the right column we report the $\nu$F$_\nu$ plots. The upper panels are relative to the modeling with a
a blackbody plus power-law, while bottom panels report the
resonant cyclotron scattering plus a power-law model (RCS+PL).} 
\label{spectraxmmintegral2}
\end{figure}

Since these flux variations are rather small, we have chosen to model
only the \XMM\, observation closest to the \INT\, one (for
\rxs\ only one \XMM\, observation is available though). Our results
from the spectral modeling of the 1--200\,keV spectrum of
\uu, \rxs, and \kes\, are summarized in Table \ref{tablexmmintegral}
and shown in Figs. \ref{spectraxmmintegral1} and \ref{spectraxmmintegral2}.

\begin{landscape}
\begin{table*}
\setlength{\tabcolsep}{0.01in}
\caption{Spectral Parameters: \uu, \rxs, \kes, and SGR 1900+14}
\begin{tabular}{lcccccccc}
\hline
\hline
 {\em Source Name} & \multicolumn{2}{c}{4U\,0142+614$^*$} &
 \multicolumn{2}{c}{1RXS\,J1708--4009$^*$} &
 \multicolumn{2}{c}{1E\,1841--045} &
 \multicolumn{2}{c}{SGR\,1900+14} \\
 
\hline
 \multicolumn{1}{l}{Parameters}
& BB+2PL & RCS+PL
& BB+2PL & RCS+PL
& BB+PL & RCS+PL
& BB+PL & RCS+PL \\
\hline
  N$_{H}$
& 1.67$^{+0.02}_{-0.02}$ & 0.81$^{+0.05}_{-0.05}$
&  1.91$^{+0.06}_{-0.06}$ &  1.67$^{+0.05}_{-0.05}$
& 2.38$^{+0.4}_{-0.1}$ & 2.57$^{+0.13}_{-0.15}$  
& 3.5$^{+0.1}_{-0.1}$ & 4.0$^{+0.1}_{-0.1}$ \\

constant
& 1.01 & 1.10
& 1.05 & 0.80
& 1.02  & 1.09
& 1.20 & 1.10 \\
 & & & & & & & &\\

kT (keV)
& 0.43$^{+0.03}_{-0.03}$ & 0.30$^{+0.05}_{-0.05}$
& 0.47$^{+0.01}_{-0.01}$ & 0.32$^{+0.05}_{-0.05}$
& 0.51$^{+0.03}_{-0.02}$ &  0.39$^{+0.05}_{-0.05}$
& 0.45$^{+0.04}_{-0.04}$ & 0.30$^{+0.08}_{-0.1}$ \\

BB~norm
&  8.7$^{+0.4}_{-0.5}\times10^{-4}$ &
&  2.4$^{+0.1}_{-0.2}\times10^{-4}$ &
& 2.4$^{+0.6}_{-0.3}\times10^{-4}$ & 
& 6.7$^{+0.1}_{-0.1}\times10^{-5}$ &  \\

$\Gamma_1$
& 4.14$^{+0.04}_{-0.04}$ &
& 2.70$^{+0.08}_{-0.08}$ &
& 
&\\

PL$_1$~norm
& 0.30$^{+0.08}_{-0.08}$ &
&  0.016$^{+0.003}_{-0.004}$ &
&  &  
&  & \\ 
 & & & & & & & & \\

$\beta_{T}$
&  & 0.33$^{+0.05}_{-0.05}$
& & 0.38$^{+0.03}_{-0.03}$
& & 0.23$^{+0.05}_{-0.05}$ 
& & 0.26$^{+0.03}_{-0.03}$ \\

$\tau_{res}$ &  & 1.9$^{+0.2}_{-0.2}$
& & 2.1$^{+0.2}_{-0.2}$
& & 1.13$^{+0.3}_{-0.2}$  
& &  2.5$^{+0.5}_{-0.2}$ \\

RCS~norm
& &  4.5$^{+0.6}_{-0.8}\times10^{-3}$
&  &  8.1$^{+1.1}_{-1.3}\times10^{-4}$
& & 3.1$^{+2.3}_{-1.1}\times10^{-4}$
& &  1.8$^{+0.04}_{-0.05}\times10^{-4}$ \\
& & & & & & & &  \\

$\Gamma_2$
& 0.78$^{+0.1}_{-0.07}$ &  1.1$^{+0.1}_{-0.1}$
& 0.76$^{+0.1}_{-0.1}$ &  1.0$^{+0.1}_{-0.1}$
& 1.47$^{+0.04}_{-0.05}$ & 1.47$^{+0.05}_{-0.05}$ 
& 1.4$^{+0.1}_{-0.1}$  & 1.24$^{+0.07}_{-0.07}$  \\

PL$_2$~norm
& 1.4$^{+0.1}_{-0.1}\times10^{-4}$ & 5.0$^{+0.1}_{-0.1}\times10^{-4}$
& 8.6$^{+0.1}_{-0.1}\times10^{-5}$ & 4.2$^{+0.1}_{-0.1}\times10^{-4}$
& 2.4$^{+0.6}_{-0.6}\times10^{-3}$& 2.2$^{+0.1}_{-0.1}\times10^{-3}$
& 4.4$^{+0.1}_{-0.1}\times10^{-4}$ & 3.0$^{+0.1}_{-0.1}\times10^{-4}$\\
& & & & & & & & \\

Flux 1--10\,keV
& 1.1$^{+0.8}_{-0.8}\times10^{-10}$ & 1.1$^{+0.8}_{-0.8}\times10^{-10}$
&2.6$^{+0.3}_{-0.3}\times10^{-11}$ & 2.6$^{+1.1}_{-0.8}\times10^{-11}$
& 2.2$^{+0.2}_{-0.3}\times10^{-11}$ & 2.1$^{+0.2}_{-0.3}\times10^{-11}$ 
& 3.9$^{+0.1}_{-0.1}\times10^{-12}$ & 3.8$^{+0.1}_{-0.1}\times10^{-12}$  \\
Flux 1--200\,keV
&2.3$^{+1.7}_{-1.1}\times10^{-10}$ & 2.3$^{+1.0}_{-1.3}\times10^{-10}$
&  1.1$^{+0.5}_{-0.5}\times10^{-10}$ & 1.4$^{+0.8}_{-0.8}\times10^{-10}$
& 1.1$^{+0.8}_{-0.8}\times10^{-10}$ & 1.1$^{+0.8}_{-0.6}\times10^{-10}$  
& 1.7$^{+0.1}_{-0.1}\times10^{-11}$ & 1.7$^{+0.1}_{-0.1}\times10^{-11}$  \\
& & & & & & & & \\

$\chi^2_{\nu}$ (dof)
&  0.99 (216) & 0.80 (216)
& 1.11 (202) & 1.01 (202)
& 1.14 (158) & 1.08 (156) 
& 1.18 (141)  &   1.15 (139)  \\

\hline
\hline
\end{tabular}
\\
\small{\\Best fit values of the spectral parameters
obtained by fitting the $\sim$1--200\,keV \XMM\, and \INT\, AXPs'
spectra with a blackbody plus two power-laws model (BB+2PL) for
\uu\, and \rxs, while a single power-law was used for \kes and SGR 1900+14. Furthermore, all the
sources were modeled with a resonant cyclotron scattering model plus a
power-law (RCS+PL). Errors are at 1$\sigma$ confidence level, reported
fluxes are absorbed and in units of \ergscm2 , and N$_{H}$ in units of
$10^{22}$\,cm$^{-2}$ and assuming solar abundances from\cite{lodders03}; 
2\% systematic error has been included. See also
Figs.\,\ref{spectraxmmintegral1}, \ref{spectraxmmintegral2}, \ref{spectra1900} and
\S\,\ref{res:hard} for details. $^{*}$: source slightly variable in
flux and spectrum, see text for details.}
\label{tablexmmintegral}
\end{table*}
\end{landscape}

In all cases we found that $N_H$, as derived from the RCS model, is
lower than (or consistent with) that inferred from the BB+2PL fit (or
BB+PL in the case of \kes), and consistent with what derived from
fitting the single X-ray edges of \uu, and \rxs \cite{durant}. This is not surprising, since the power-law usually
fitted to magnetar spectra in the soft X-ray range is well known to
cause an overestimate in the column density\footnote{This is because
the absorption model tends to increase the $N_H$ value in response of
the steep rise of the power-law at low energies, which eventually
diverges approaching E=0.} The surface temperature we derived fitting
the RCS model is systematically lower than the corresponding BB
temperature in the BB+2PL or BB+PL models, and is consistent with
being the same ($\sim 0.33$\,keV) in the four sources. On the other
hand the thermal electron velocity and the optical depth are in the
ranges 0.2--0.4 and 1.0--2.1, respectively. Concerning the hard X-ray
power-law, we find that the photon index is, within the errors, the
same when fitting the RCS or the BB+2PL or BB+PL models, while the hard PL normalization is larger in
the RCS case with respect to the BB+2PL model. Both the soft and the
hard X-ray fluxes of all these AXPs derived from the RCS fitting are
consistent with those implied by the usual BB+2PL fitting.

\subsection{SGRs}
\label{res:sgrs}

\begin{figure*}
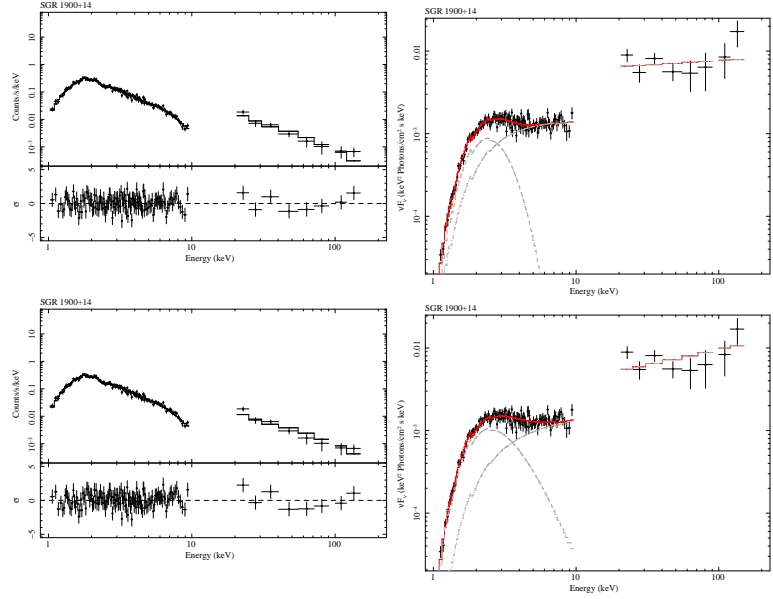

\centering{
\hspace{0.1cm}
\vbox{
\hbox{
\includegraphics[height=2.cm,angle=270,width=5.cm]{f9a.eps}
\includegraphics[height=2.cm,angle=270,width=5.cm]{f9b.eps}}
\hbox{
\includegraphics[height=2.cm,angle=270,width=5.cm]{f9c.eps}
\includegraphics[height=2.cm,angle=270,width=5.cm]{f9d.eps}}}}
\caption{\sgrb: left column shows the spectra in Counts/s/keV while in the right
column we report the $\nu$F$_\nu$ plots. The upper panels are relative
to the modeling with a blackbody plus power-law (BB+PL), while bottom
panels report the resonant cyclotron scattering model plus power-law
(RCS+PL). 
The red lines represent the total model, while the dashed
lines are the single components.}
\label{spectra1900}
\end{figure*}

Finally, we consider the 1--200\,keV emission of SGRs
(see Table\,\ref{tablexmmintegral} and Fig.\,\ref{spectra1900}).
It has been already noticed that the hard X-ray emission of SGRs is
quite different from that of AXPs \cite{dg06}. In fact, the spectra of AXPs show a clear turnover between 10 and 20
keV (see Fig. \ref{spectraxmmintegral1} and \ref{spectraxmmintegral2}) and the fit requires an
additional spectral component. Instead, the hard X-ray emission of
SGRs seems the natural continuation of the non-thermal component which
is dominant in the 1--10 keV energy range. This is why we can use a BB
(or RCS) plus a single power-law in the entire 1--200\,keV range, 
while for the hard X-ray emitting AXPs we were forced to add a
second power-law to the BB+PL model.

Despite the fact that \sgra\ and \sgrb\ are both detected by \INT, here we report 
the results on \sgrb\ only, since \sgra\ has been proven to be variable an time scales
which are shorter than the ones required by \INT/IBIS in order to obtain a good quality
spectrum, and hence we are not able to deal with a truly simultaneous broad band spectrum.


In the \sgrb\, 1--200\,keV spectrum, we found consistent $N_H$ and
spectral index values between the BB+PL and RCS+PL models, and a RCS
surface temperature significantly lower than the corresponding BB
temperature. In all the SGR observations, the derived fluxes are
consistent among the two models.

\section{Discussion}
\label{discussion}

Before discussing our results and the physics we can derive from our
model, we would like to stress once again that the RCS model involves
a number of simplifications (see \S\ref{rcsmodel}). One is the
assumption of a single temperature surface emission. Current-carrying
charges will hit and heat the star surface, generally inhomogeneously
(TLK). In addition, the emission emerging from the surface is likely
to be non-Planckian. While the presence of an atmosphere on top the
crust of a magnetar remains a possibility \cite{guver07a,guver07b}, its properties, are then likely different from those of
a standard (in radiative and hydrostatic equilibrium) atmosphere on,
e.g., a canonical isolated cooling neutron star (see e.g. \cite{ho03}). 
The extreme field and (relatively)
low surface temperature ($\lesssim 0.5\ {\rm keV}$) of magnetar
candidates may also be suggestive of a condensed surface, at least if
the chemical composition is mainly Fe \cite{tzd04}. 
In the light of these considerations, and in the absence of a
detailed model for the surface emission, and for the atmosphere of
strongly magnetized NSs constantly hit by returning currents, we
restricted ourself to a blackbody approximation for the seed thermal
photons.

In spite of these simplifications, \cite{rea08} found that the RCS model can
describe the soft X-ray portion of the whole set of magnetar spectra
we have considered there, including the TAXPs variability, by using only
three free parameters (plus a normalization factor). This is the same
number of degrees of freedom required by the blackbody plus power law
model, commonly used to fit this energy band.

\subsection{Magnetar magnetospheric properties}

One of the most interesting outcomes of our analysis is the
measure of the magnetospheric properties of magnetars.  In all
sources, steady and variable ones (see \cite{rea08}), the value of $\tau_{res}$ is in the
range of $\sim 1$--$6$. This suggests that the entire
class of sources are characterized by similar properties of scattering
electrons, their density and their (thermal) velocity spread. An
optical depth $\tau_0=\tau_{res}/2$ requires a particle density $n_e$
(see eq.~[\ref{tau0}]) which can be easily inferred considering:
\begin{equation} \tau_0 \approx 1.8
\times 10^{-20} n_e r_{sc} \left(\frac{ 1\, {\rm keV}}{ \hbar
\omega_B}\right)  \, , \label{tau00}
\end{equation}
where $r_{sc}$ is the
radius of the scattering sphere
\begin{equation} r_{sc} \approx 8
R_{NS}\left (\frac{B}{B_{crit}}\right )^{1/3} \left (\frac{ 1\, {\rm
keV}}{ \hbar \omega_B} \right )^{1/3} \, , \label{rsc}
\end{equation}
$R_{NS}$ is the neutron star radius and $B_{crit} \approx 4.4 \times
10^{13}$~G is the quantum critical field.  By taking a typical photon
energy of $\sim 1\ {\rm keV}$, $R_{NS} \sim 10^6$~cm and $B\sim 10
B_{crit}$, we get $n_e \approx 1.5\times 10^{13} \tau_{res}\, {\rm
cm}^{-3}$.  This is several orders of magnitude larger than the
Goldreich-Julian density \cite{Goldandjul69} at the same
distance, $n_{GJ} \approx n_e \pi r_{sc} /(3
\tau_{res} R_{lc}) \sim 2 \times 10^{10}$~cm$^{-3}$ (where $R_{lc}$ is the light
cylinder radius and we took $P\sim 10$~s). While the charge density is
large when compared with the minimal Goldreich-Julian density, it
provides a negligible optical depth to non-resonant Thomson
scattering. Only the resonant cyclotron scattering makes an efficient
photon boosting possible.

Our present model does not include a proper treatment of
magnetospheric currents, so that $\tau_{res}$ is a free parameter
related to the electron density. Nevertheless, it is useful to compare
the values of the optical depth inferred here to those expected when a
current flow arises because a steady twist is implanted in the star
magnetosphere, as in the case investigated by TLK under the assumption
of axisymmetry and self-similarity. If the scattering particles
have a collective motion (bulk velocity $\beta_{bulk}$), the
efficiency of the scattering process is related to $\tau_{res}
\beta_{bulk}$ (e.g. \cite{ntz93}). This quantity
is shown as a function of the magnetic colatitude in Fig.~5 of TLK for
different values of the twist angle, $\Delta\phi_{N-S}$. By assuming
$\beta_{bulk}=1$ and integrating over the angle, we get the average
value of the scattering depth as a function of $\Delta\phi_{N-S}$.
A value of $\tau_{res} \sim 1$ is only compatible with very large values of the twist angle
(i.e $\Delta\phi_{N-S} > 3$), while typical values of $\tau_{res} \sim
2$, as those obtained from some of our fits, require $\beta_{bulk}
\lesssim 0.5$ to be compatible with $\Delta\phi_{N-S}\sim 3$ (the
smaller is $\beta_{bulk}$, the smaller is the value of the twist
angle). This is consistent with the fact that the RCS model has been
computed under the assumption of vanishing bulk velocity for the
magnetospheric currents, and it is compatible with TLK model only when
in the latter it is $\beta_{bulk}\ll 1$.

\section{Conclusions}

In the last few years the detection of bursts from AXPs strengthened their connection with
SGRs. However, the latter behave differently in many respects.  Below
$\sim4$~keV, the SGRs emission can be described either by a blackbody
or an RCS component. At higher energies though ($>4$\,keV), their
spectra require the addition of a power-law component, which well
describes the spectrum until $\sim 200$\,keV. The non-thermal
component dominates their spectra to the point that the choice of a
blackbody or the RCS model at lower energies does not affect
significantly the value of the hard X-ray power-law index, nor the
energy at which this component starts to dominate the spectrum. 
The spectra of SGRs are then strongly
non-thermally dominated in the 4--200\,keV range.

The case of the AXPs is different (with the exception of
\kes). These sources show a more complex spectrum, with an
evident non-thermal component below $\sim10$~keV, apparently different
from that observed at higher energies. For the AXPs detected at
energies $>$20\,keV, the spectrum can be described by a RCS component
until 5--8\,keV, above which the non-thermal hard X-ray component
becomes important, and (e.g. for \rxs\, and \uu) dominates until $\sim
200$\,keV. In the case of the BB+2PL model instead, the non-thermal
component responsible for the hard X-ray part of the spectrum starts
to dominate only above $\sim 10$~keV (see e.g.
Figs. \ref{spectraxmmintegral1} and \ref{spectraxmmintegral2}).   
It is worth noting that the photon index of the hard X-ray
component in AXPs does not strongly depend on the modeling of the
spectrum below 10\,keV, while, its normalization and, as a
consequence, the value at which the hard tail starts to dominate the
spectrum, do.

The fact that hard X-ray spectra detected from AXPs are much flatter
than those of SGRs may also suggest a possible difference in the
physical mechanism that powers the hard tail in the two classes of
sources. Within the magnetar scenario, \cite{thompson05}
discussed how soft $\gamma$-rays may be produced in a twisted
magnetosphere, proposing two different pictures: either thermal
bremsstrahlung emission from the surface region heated by returning
currents, or synchrotron emission from pairs created higher up ($\sim$
100 km) in the magnetosphere. Moreover, a third scenario involving
resonant magnetic Compton up-scattering of soft X-ray photons by a
non-thermal population of highly relativistic electrons has been
proposed by \cite{baring07}. It is interesting to note that
3D Monte Carlo simulations \cite{fernandez07,ntz08} show that multiple peaks may appear in the
spectrum. In particular, in the model by \cite{ntz08}, a second ``hump'' may be present when up-scattering is so
efficient that photons start to fill the Wien peak at the typical
energy of the scattering electrons. The change in the spectral slope
may be due, in this scenario, to the peculiar, ``double-humped'' shape
of the continuum. The precise localization of a possible down-break at higher energies is
therefore of great potential importance and might provide useful
information on the underlying physical mechanism responsible for the
hard emission.

\acknowledgments{D. G. aknowledges the French Space Agency (CNES) for financial support.  N.R. is supported by an NWO Veni Fellowship, and acknowledges the warm hospitality of the Mullard Space Science Laboratory, where this work was started, and of the Purdue University where it has been completed. S.Z. acknowledges STFC for support through an Advanced Fellowship. Based on observations with {\it INTEGRAL} and {\it XMM-Newton}, two ESA missions with instruments and science data centres funded by ESA member states with the participation of Russia and the USA.
The RCS model is available to the community on the {\tt XSPEC}
website\footnote{http://heasarc.gsfc.nasa.gov/docs/xanadu/xspec/models/rcs.html}.}

\end{document}